\documentclass[sigconf]{acmart}

\usepackage{amsmath}
\usepackage{booktabs}
\usepackage{subcaption}
\usepackage{balance}
\graphicspath{{figures/}}

\copyrightyear{2026}
\acmYear{2026}
\setcopyright{cc}
\setcctype{by}
\acmConference[SpatialConnect '26]{The 2nd ACM SIGSPATIAL International Workshop on Spatial Intelligence for Smart and Connected Communities}{November 03--06, 2026}{Riverside, CA, USA}
\acmBooktitle{The 2nd ACM SIGSPATIAL International Workshop on Spatial Intelligence for Smart and Connected Communities (SpatialConnect '26), November 03--06, 2026, Riverside, CA, USA}
\acmDOI{10.1145/3849741.3856137}
\acmISBN{979-8-4007-3063-4/2026/11}

\begin{document}

\title[Measuring Rural Healthcare Accessibility and Resilience]{Measuring Healthcare Accessibility and Resilience for Smart and Connected Rural Communities: A Florida Panhandle Case Study}


\author{Dahai Yu}
\email{dahai.yu@fsu.edu}
\affiliation{%
  \institution{Florida State University}
  \city{Tallahassee}
  \state{Florida}
  \country{USA}
}




\author{Zhe He}
\email{zhe@fsu.edu}
\affiliation{%
  \institution{Florida State University}
  \city{Tallahassee}
  \state{Florida}
  \country{USA}
}

\author{Amber DeJohn}
\email{amber.dejohn@fsu.edu}
\affiliation{%
  \institution{Florida State University}
  \city{Tallahassee}
  \state{Florida}
  \country{USA}
}

\author{Xinyue Ye}
\email{xye10@ua.edu}
\affiliation{%
  \institution{University of Alabama}
  \city{Tuscaloosa}
  \state{Alabama}
  \country{USA}
}

\author{Guang Wang}
\authornote{Prof. Guang Wang is the corresponding author.}
\email{guang@cs.fsu.edu}
\orcid{0000-0002-7739-7945}
\affiliation{%
  \institution{Florida State University}
  \city{Tallahassee}
  \state{Florida}
  \country{USA}
}

\renewcommand{\shortauthors}{Yu et al.}

\begin{abstract}
Rural communities in the United States face persistent healthcare disparities, with fewer providers and facilities, longer trips to care, lower health literacy, and weaker transportation and broadband infrastructure than their non-rural counterparts. Beyond any single barrier, healthcare accessibility is shaped jointly by the geographic availability of services, residents' ability to reach those services, realized utilization, and the capacity of local systems to remain operational during disruptions, which are often examined separately and at coarse spatial scales, limiting their usefulness for community planning. We present a fine-grained, multi-source longitudinal measurement study of healthcare accessibility in Florida, with a focus on the hurricane-prone Florida Panhandle. We integrate healthcare-facility points of service, monthly mobility records from January 2018 through April 2021, Census Block Group (CBG)-level demographic data, and road-network data to compare rural and non-rural communities from supply, travel, utilization, and resilience perspectives. We find that approximately 95.5\% of the Panhandle's land area is rural and that 46.7\% of rural CBGs contain no healthcare facility in the pooled inventory. Rural residents have less than half the per-capita facility availability of non-rural residents (1.50 versus 3.75 facilities per 1,000 residents). For origin--facility pairs represented in the mobility panel, rural road distance averages 20.5 miles, compared with 9.2 miles for non-rural origins; the corresponding modeled travel times are 26.4 and 11.8 minutes. Population density explains much of this raw travel difference, although a smaller rural association remains after measured adjustment. Longitudinal results show disruptions in observed healthcare activity around Hurricane Michael and COVID-19. A transparent access screen identifies 45 rural CBGs containing 53,612 residents with low potential access, no local ambulatory facility, and more than 15 minutes of modeled travel to the nearest facility. We connect these measured priorities to planning options for mobile clinics, non-emergency medical transportation, telehealth, and disaster-resilient services. The study demonstrates how community-oriented spatial intelligence can support equitable and resilient healthcare planning in rural regions.
\end{abstract}



\maketitle

\section{Introduction}

Access to healthcare services plays a pivotal role in improving health outcomes, yet rural residents face a variety of access barriers, such as a shortage of healthcare professionals and facilities, long distances to healthcare services, lack of reliable transportation, lower health literacy, and limited broadband access~\cite{douthit2015exposing,hartley2004rural, yu2026healthmamba}. A shortage of healthcare professionals or facilities limits the availability of timely and specialized care, while long travel distances to medical facilities create additional challenges, especially for individuals with chronic conditions or limited mobility~\cite{syed2013traveling,weiss2020globalhealthcare}. Many rural residents also lack reliable transportation, further compounding the difficulty of reaching care when it is needed most~\cite{wallace2005access}. In addition, lower levels of health literacy can hinder patients' ability to understand, interpret, and act on medical information, resulting in delayed treatment or poor adherence to care plans~\cite{berkman2011low}. Limited broadband access presents another obstacle, as it restricts the use of telemedicine and digital health tools that could otherwise bridge geographical gaps~\cite{hirko2020telehealth}. Together, these structural and informational barriers exacerbate health access gaps between rural and non-rural populations, underscoring the urgent need for innovative, community-centered solutions to improve healthcare access in rural areas.

Importantly, healthcare access is not determined solely by whether a facility exists. It also depends on whether an appropriate service is available within a feasible travel range, whether residents have the transportation and information needed to obtain care, and whether the local healthcare system remains operational during disasters~\cite{hong2021resilience}. These conditions are especially difficult to satisfy in rural communities, where the barriers above reinforce one another and interact with low population density and an aging population. The resulting access gap can delay preventive care, complicate chronic-disease management, and increase the burden on residents who are least able to absorb additional travel time and cost.

The sustainability of rural healthcare provision is similarly fragile. Rural hospital inpatient volumes have declined in recent years, driven in part by rural Medicare beneficiaries' increasing use of urban hospitals~\cite{friedman2022rural}. Lower patient volumes and weaker financial performance have also been associated with subsequent rural hospital closure~\cite{kozhimannil2018obstetric}. During hurricanes, pandemics, and other emergencies, rural systems face a compound challenge: facilities, roads, power, communications, and staffing may be disrupted just as healthcare needs increase~\cite{hong2021resilience}. A planning framework for rural healthcare must therefore examine \textit{routine accessibility} and \textit{disruption resilience} together, and support both \textit{\textbf{patient access}} and \textit{\textbf{provider sustainability}}, rather than treating them as separate problems.

Current approaches to rural healthcare disparities typically revolve around expanding digital health initiatives and subsidizing rural clinics. However, without reliable connectivity, state-of-the-art telehealth solutions remain out of reach for many vulnerable rural populations~\cite{hirko2020telehealth}, necessitating a broader, more integrated approach that combines physical transportation solutions, educational interventions, and infrastructural resilience. Designing such an approach requires a comprehensive empirical understanding of where the gaps in the healthcare network lie, which services are most needed, and how access and utilization respond to disruptions.

Spatial intelligence provides an opportunity to build this understanding. Facility inventories characterize potential supply; mobility-derived visitation patterns reveal realized interactions between communities and providers; demographic data identify populations for whom an equivalent spatial barrier may have more severe consequences; and longitudinal observations show how the system responds to shocks. However, prior studies often focus on one dimension, e.g., provider-to-population ratios, distance to the nearest facility, or aggregate visit counts, and many concentrate on metropolitan areas. A boutique rural analysis requires a more integrated view at a sufficiently fine geographic scale.

In this work, we conduct a data-driven longitudinal analysis using the Florida Panhandle as a representative case study. The Panhandle is a compelling test bed because it combines extensive rural territory, an aging population, limited service availability, and recurring hurricane exposure---conditions that jointly stress-test rural healthcare systems. We integrate healthcare-facility data, monthly mobility records from January 2018 through April 2021, CBG-level demographic information, and road-network data. Our analysis is organized around four research questions:

\begin{description}
    \item[RQ1: Spatial supply.] How unevenly are healthcare facilities distributed between rural and non-rural communities?
    \item[RQ2: Access burden.] How do observed travel distance and age-related healthcare needs differ between rural and non-rural residents?
    \item[RQ3: Temporal dynamics and hazard resilience.] How do facility availability and utilization evolve over time, particularly around major hurricanes and the COVID-19 pandemic?
    \item[RQ4: Community actionability.] How can spatial evidence be operationalized into equitable intervention prioritization, deployment, and evaluation decisions?
\end{description}

The key contributions of this work are as follows:
\begin{itemize}
    \item We conduct a fine-grained multidimensional characterization of rural healthcare accessibility that jointly considers potential supply, observed travel, realized utilization, demographic need, and disruption dynamics using real-world large-scale data.
    \item We quantify substantial rural--non-rural disparities in the Florida Panhandle and test their stability through time-aligned facility definitions, population-density and demographic adjustment, road-network travel, and county-block confidence intervals.
    \item We analyze how facility presence and observed activity change during public emergencies, while carefully distinguishing potential access, destination visits, and panel-observed visitors by origin.
    \item We apply a transparent community access screen and connect the resulting Panhandle priorities to E2SFCA-based planning, stakeholder input, constrained intervention design, and equity-aware evaluation.
\end{itemize}

\section{Related Work}
\subsection{Rural Healthcare Access and Sustainability}
A broad public-health literature documents structural barriers to rural healthcare, including provider shortages, transportation constraints, limited broadband, and reduced access to specialized care~\cite{douthit2015exposing}. Transportation barriers are associated with missed appointments, delayed care, and disrupted chronic-disease management~\cite{syed2013traveling}; non-emergency medical transportation (NEMT) has therefore become an important intervention for residents who cannot reliably reach care~\cite{wallace2005access}. Telehealth can reduce some physical travel, but inadequate connectivity and digital readiness can prevent rural populations from benefiting equally~\cite{hirko2020telehealth}. Rural healthcare supply also faces financial pressure: lower patient volumes and broader market conditions have been associated with increasing rural hospital closures~\cite{kaufman2016rising}. Our work links these well-established barriers to fine-scale spatial evidence within one regional system.

\begin{table*}[t]
\centering \footnotesize
\caption{Data layers and their descriptions.}
\label{tab:data}
\begin{tabular}{p{0.15\textwidth}p{0.30\textwidth}p{0.36\textwidth}p{0.1\textwidth}}
\toprule
Layer & Spatial/temporal resolution & Main variables & Source \\
\midrule
Healthcare facilities & Point locations; pooled inventory and monthly records & Facility identity, coordinates, NAICS category & SafeGraph and Advan \\
Mobility & POI-month totals and home-CBG visitor counts & Visits to POIs and panel-observed visitors by origin & SafeGraph and Advan \\
Demographics and roads & 2018 ACS 5-year estimates; 2018 road segments & Population, socioeconomic covariates, road class & US Census\\
\bottomrule
\end{tabular}
\end{table*}

\subsection{Spatial Measurement of Healthcare Accessibility}
Healthcare access is multidimensional. The classic Penchansky--Thomas framework identifies availability, geographic accessibility, affordability, awareness, and acceptability~\cite{saurman2016improving}. Our empirical analysis focuses primarily on availability and geographic accessibility, leaving awareness and other dimensions to future work. Place-based accessibility methods estimate the opportunities reachable from a geographic location given the spatial distribution of destinations and transportation constraints~\cite{higgins2022calculating,weiss2020globalhealthcare,yu2025uqgnn}. In healthcare, GIS-based studies commonly operationalize this concept using provider-to-population ratios, nearest-facility distance, or two-step floating catchment area (2SFCA) methods~\cite{mcgrail2012spatial}. The 2SFCA family is particularly valuable because it accounts for competition for services across administrative boundaries and can incorporate distance decay and rural--urban differences in catchment size. These place-based accessibility methods generally estimate \emph{potential} access, i.e., what residents could reach given the spatial distribution of supply and demand. Mobility-derived visits complement this view with \emph{realized} access: where residents actually travel and which facilities they use in practice. Our analysis combines both perspectives, pairing the structural view from 2SFCA-style measures with the behavioral signal captured in observed visitation patterns.

\subsection{Mobility Data for Public-Health Analysis}
Aggregated mobility and POI data have supported public-health studies ranging from epidemic modeling to inequality measurement~\cite{chang2021mobility,jay2020income,kang2020multiscale,grantz2020mobile}, offering a scale and temporal resolution that traditional survey-based approaches cannot easily match~\cite{gonzalez2008mobility,alessandretti2020scales,schlapfer2021visitation}. Building on this line of work, we treat mobility-derived activity as a measure of observed healthcare-seeking behavior within the data provider's panel. Because smartphone-derived panels can vary across geography, urbanicity, and demographic groups~\cite{li2023bias,brelsford2022publicspace,grantz2020mobile, yu2026energymamba}, we complement raw counts with population-based normalization. This normalization supports comparison across communities of different population sizes.

\subsection{Spatial Intelligence for Smart and Connected Communities}
Smart and connected community (S\&CC) research seeks not only to describe disparities but also to turn data into actionable improvements in public services and infrastructure. For rural healthcare, this requires connecting diagnostics to decisions such as where to locate mobile services, how to route NEMT vehicles, which communities should receive broadband and telehealth investments first, and how to preserve access during disasters. Related work on post-disaster power restoration shows that coupling uncertainty-aware prediction with equity-aware optimization can shorten outages while reducing disparities between communities~\cite{jiang2025epopr}. Our study contributes both an evidence layer that quantifies where and how access gaps arise and a deployment-oriented workflow that translates this evidence into prioritized, actionable interventions.

\section{Study Context and Community Relevance}
\label{sec:context}
\subsection{Florida Panhandle}
Our statewide data provide geographic context, while the empirical analyses focus on the Florida Panhandle, the northwestern portion of the state comprising 18 counties. The region contains large rural areas, dispersed settlements, and communities exposed to hurricanes and other coastal hazards. Under the CBG-level classification used in our empirical analysis, approximately 95.5\% of the Panhandle's land area and 37.7\% of its population are rural.



\subsection{Stakeholder-Defined Community Challenge}
The empirical problem addressed here is the difficulty of deciding \emph{where} limited rural-health resources should be deployed and \emph{which} intervention is appropriate for each community. A CBG with no nearby ambulatory care may require a different response from a CBG whose primary barrier is transportation, broadband, or disaster vulnerability.

\section{Data and Methods}
\subsection{Data Sources}
Table~\ref{tab:data} summarizes the three data sources used for this study. All geographic data are harmonized to a common coordinate reference system and linked through facility coordinates, CBG boundaries, and origin CBG identifiers.

 \begin{figure}[t]
  \centering
  \includegraphics[width=\linewidth]{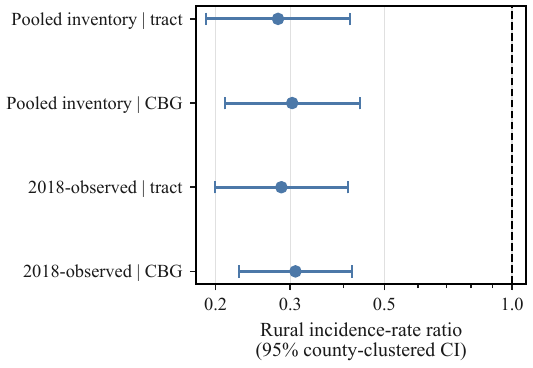}
  \Description{Four interval estimates show rural incidence-rate ratios below one in CBG and tract negative-binomial models.}
  \caption{Adjusted negative-binomial rural IRRs for facility supply. The pooled master is an inventory across product vintages. CBG and tract count models use population offsets, county fixed effects, and county-clustered standard errors.}
  \label{fig:supply_adjusted}
\end{figure}

\textbf{Healthcare facilities and visits.} We use a large-scale POI dataset of healthcare-related facilities in Florida and associated monthly mobility records from January 2018 through April 2021 collected by SafeGraph/Advan~\cite{safegraph2022global, safegraph2025monthly}. 
Each facility is spatially joined to a CBG and labeled according to that CBG's rural--non-rural status. The POI master pools multiple product vintages, so we treat it as an inventory and repeat the supply analysis using only POIs observed in at least one 2018 Patterns record. The Patterns field \texttt{raw\_visit\_counts} measures visits to destination POIs, whereas \texttt{visitor\_home\_cbgs} contains privacy-processed counts of panel visitors by home CBG. We therefore refer to origin-specific quantities as \emph{panel-observed visitors}, rather than visits or unique residents.

\textbf{Demographic and road data.} We obtain population, median age, household income, uninsured share, race and ethnicity, educational attainment, and broadband availability from the 2014--2018 American Community Survey (ACS) 5-year Summary Files~\cite{census2018acs}. Because detailed disability, poverty, and household vehicle tables are not released at CBG level for this vintage, tract-level sensitivity models include these additional variables. Educational attainment is used only as a contextual correlate of potential health-literacy barriers~\cite{kutner2006health}, not as a direct measure of individual health literacy. We construct a motor-vehicle network from 2018 county TIGER/Line road files~\cite{census2018tiger}. CBGs are classified as urban when at least 34.5\% of their land area overlaps a 2010 Census Urban Area and as rural otherwise~\cite{census2010urban}. The cutoff was selected in the original statewide data preparation by comparing candidate cutoffs from 20\% to 49.5\% and selecting the one for which the total area of whole CBGs labeled urban most closely matched the area of the Census Urban Area polygons. This calibration preserves the statewide amount of urban land when boundary-crossing CBGs must receive a single label; it is not a natural divide between two kinds of communities. Table~\ref{tab:rurality_sensitivity} therefore repeats the main supply and travel summaries at 25\% and 50\%: the number of rural CBGs changes, but rural facility availability remains about 36.5\%--40.0\% of non-rural availability, and rural modeled driving time remains a little more than twice the non-rural mean.

\begin{table}[t]
\centering
\small
\setlength{\tabcolsep}{3pt}
\caption{Sensitivity to the CBG land-overlap cutoff. Rate ratio: rural facilities per 1,000 residents divided by the non-rural rate. Time ratio: rural visitor-weighted mean driving time divided by the non-rural mean.}
\label{tab:rurality_sensitivity}
\begin{tabular}{rrrrrr}
\toprule
Cutoff & \shortstack{Rural\\CBGs} & \shortstack{Rural\\pop.} & \shortstack{Rural\\land} & \shortstack{Rate\\ratio} & \shortstack{Time\\ratio} \\
\midrule
25.0\% & 341 & 34.64\% & 94.80\% & 0.365 & 2.251 \\
34.5\% & 364 & 37.70\% & 95.50\% & 0.400 & 2.234 \\
50.0\% & 408 & 43.39\% & 96.39\% & 0.381 & 2.175 \\
\bottomrule
\end{tabular}
\end{table}


\textbf{Disaster events.} Hurricane Michael made landfall in the Florida Panhandle in October 2018 and severely affected Panhandle communities~\cite{nhc2019michael,hong2021resilience}. Bay County is the focal case. Escambia and Santa Rosa Counties provide western comparisons and were outside the 12 counties designated for FEMA Individual Assistance under DR-4399-FL~\cite{florida2018michael}. We also examine the period around the national onset of COVID-19 disruptions in March 2020.

\subsection{Accessibility and Utilization Measures}
We operationalize healthcare accessibility through complementary measures rather than a single score.

\textbf{Facility availability.} For group $g\in\{\text{rural},\text{non-rural}\}$, per-capita availability is
\begin{equation}
A_g = 1000\,\frac{\sum_{i\in g} F_i}{\sum_{i\in g} P_i},
\end{equation}
where $F_i$ is the number of facilities in CBG $i$ and $P_i$ is its population. We additionally report the zero-facility rate
\begin{equation}
Z_g = \frac{\sum_{i\in g}\mathbb{I}(F_i=0)}{|g|}.
\end{equation}

\textbf{Observed travel burden.} Let $v_{ij}$ denote the panel-observed visitor count from origin CBG $i$ to facility $j$, and let $d_{ij}$ be the distance or modeled travel time between the origin and facility. The visitor-weighted mean is
\begin{equation}
D_g = \frac{\sum_{i\in g}\sum_j v_{ij}d_{ij}}{\sum_{i\in g}\sum_j v_{ij}}.
\end{equation}

Origins are represented by Census internal points and destinations by POI coordinates. We connect TIGER/Line shape vertices, retain the largest connected component (99.66\% of nodes), and snap origins and facilities to the network. Because TIGER/Line does not report speeds, modeled off-peak travel time uses stated speeds by road class; road distance provides a speed-independent comparison.


\begin{figure}[t]
  \centering
  \begin{subfigure}[t]{0.15\textwidth}
    \centering
    \includegraphics[width=\linewidth]{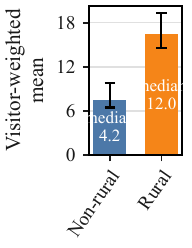}
    \Description{A paired bar chart compares visitor-weighted straight-line distance for non-rural and rural origins.}
    \caption{Straight-line distance (miles).}
  \end{subfigure}
  \hfill
  \begin{subfigure}[t]{0.15\textwidth}
    \centering
    \includegraphics[width=\linewidth]{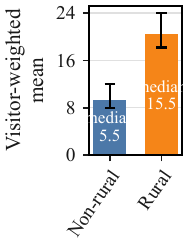}
    \Description{A paired bar chart compares visitor-weighted road-network distance for non-rural and rural origins.}
    \caption{Road-network distance (miles).}
  \end{subfigure}
  \hfill
  \begin{subfigure}[t]{0.15\textwidth}
    \centering
    \includegraphics[width=\linewidth]{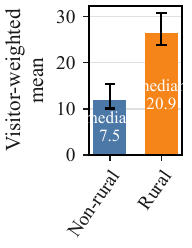}
    \Description{A paired bar chart compares visitor-weighted modeled driving time for non-rural and rural origins.}
    \caption{Modeled driving time (minutes).}
  \end{subfigure}
  \caption{2018 visitor-weighted travel burden. Bars are means and error bars are county-block 95\% bootstrap intervals; labels inside bars are weighted medians.}
  \label{fig:road_travel}
\end{figure}

\textbf{Utilization and potential access.} Monthly utilization is measured using destination visits per observed facility and origin panel visitors per 1,000 2018 ACS residents. These measures distinguish activity at rural facilities from activity among residents of rural CBGs. Because observed activity reflects population size, facility supply, and mobile-panel coverage in addition to underlying need, it is not interpreted as latent demand. We also compute an E2SFCA score with a 60-minute catchment and four decay zones: 0--15, 15--30, 30--45, and 45--60 minutes, weighted 1.00, 0.68, 0.22, and 0.05. Each POI observed in 2018 supplies one facility-equivalent because staffing and capacity are unavailable, and ACS population supplies the demand term.

\textbf{Temporal dynamics and disruption response.} We analyze monthly facility presence and visit activity from January 2018 through April 2021. For an event at month $t_0$, an interpretable relative change can be calculated as
\begin{equation}
\Delta_{g,t}=\frac{Y_{g,t}-\bar{Y}_{g,\mathrm{pre}}}{\bar{Y}_{g,\mathrm{pre}}},
\end{equation}
where $Y_{g,t}$ is a monthly outcome and $\bar{Y}_{g,\mathrm{pre}}$ is the pre-event baseline. 

\textbf{Adjustment and uncertainty.} Facility counts are modeled with a population offset, county fixed effects, and county-clustered standard errors. Negative-binomial models address marked Poisson overdispersion; CBG models use the demographic variables above, while tract-level sensitivity models add disability, poverty, household vehicle availability, and broadband. Travel models jointly include rurality, median age, the CBG controls, and county fixed effects and are weighted by panel visitors. Bar-chart intervals use county-block bootstrap resampling with 10,000 draws for supply and 2,000 draws for mobility outcomes. For Hurricane Michael, nonnegative donor weights minimize January--September 2018 pre-event error after each county series is normalized to its own pre-event mean.

\section{Results}

\subsection{RQ1: Rurality and Healthcare Supply}
As Section~\ref{sec:context} notes, the Panhandle is geographically dominated by rural territory: rural CBGs account for about 95.5\% of the land area but only 37.7\% of the population. This contrast, a large share of territory but a smaller share of population, illustrates the core service-delivery challenge: providers must cover long distances while drawing from relatively small local populations.

Rural and non-rural communities exhibit a substantial supply disparity. Non-rural areas contain 3,490 facilities, compared with 845 in rural areas. Population normalization does not remove the gap: non-rural areas have 3.75 facilities per 1,000 residents, whereas rural areas have 1.50; their county-block 95\% bootstrap intervals are 3.28--4.29 and 1.20--1.83, respectively. Restricting the inventory to POIs observed in 2018 lowers the overall rates to 1.02 and 2.59 but retains the rural--non-rural disparity. In addition, 170 of 364 rural CBGs (46.7\%) contain no inventory facility, compared with 129 of 560 non-rural CBGs (23.0\%). These CBGs are not necessarily devoid of access because residents may cross CBG boundaries, but they have no local supply and are consequently more dependent on transportation and neighboring service systems.

The disparity also remains after measured demographic adjustment (Figure~\ref{fig:supply_adjusted}). In the CBG negative-binomial model, rurality has an incidence-rate ratio (IRR) of 0.30 (95\% CI 0.21--0.44) after adjustment for income, uninsured share, race and ethnicity, county fixed effects, and population. A tract-level model that additionally includes disability, poverty, household vehicle availability, and broadband gives an IRR of 0.28 (0.19--0.41). The corresponding estimates using the 2018-observed facility definition are 0.31 (0.23--0.42) and 0.29 (0.20--0.41).

\subsection{RQ2: Travel and Demographic Burden}
\begin{figure}[t]
  \centering
  \begin{subfigure}[t]{0.73\linewidth}
    \centering
    \includegraphics[width=\linewidth]{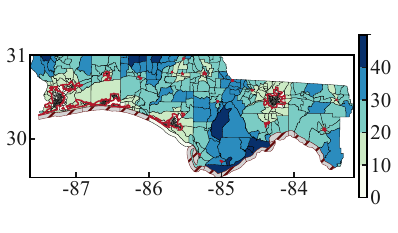}
    \Description{A Panhandle Census Block Group map shows mean straight-line distance per represented healthcare visit, with urban-area boundaries outlined in red.}
    \caption{Travel Distance (Miles) Per Visit.}
  \end{subfigure}
  \hfill
  \begin{subfigure}[t]{0.24\linewidth}
    \centering
    \includegraphics[width=\linewidth]{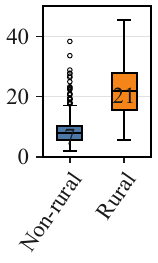}
    \Description{Box plots compare the distribution of mean straight-line distance per represented visit across rural and non-rural Panhandle Census Block Groups.}
    \caption{Travel Distance Per Visit.}
  \end{subfigure}
  \caption{Observed travel distance to healthcare facilities for rural and non-rural origins. Panel (a) maps each CBG's mean straight-line distance per represented visit; panel (b) compares the unweighted CBG distributions.}
  \label{fig:travel_distance}
\end{figure}

Observed healthcare trips originating in rural communities are substantially longer (Figure~\ref{fig:road_travel}). Across the 122,246 origin--facility pairs represented in the mobility panel, rural origins have a visitor-weighted mean straight-line distance of 16.5 miles, compared with 7.5 miles for non-rural origins. Road routing increases these means to 20.5 and 9.2 miles, and modeled travel times average 26.4 and 11.8 minutes. The corresponding weighted medians are 12.0 versus 4.2 miles by straight line, 15.5 versus 5.5 miles by road, and 20.9 versus 7.5 minutes. County-block 95\% bootstrap intervals preserve the same separation across all three measures. This difference implies additional fuel cost, travel time, scheduling difficulty, and time away from work or caregiving. Figure~\ref{fig:travel_distance} complements these visitor-weighted trip statistics by mapping each CBG's mean straight-line distance per represented visit and showing the corresponding unweighted CBG distributions.

Population density accounts for much of the raw contrast (Table~\ref{tab:travel_decomposition}). The staged models compare rural and non-rural origins within the same county, then add population density and the CBG-level demographic measures, and finally add E2SFCA potential access and the tract-level share of households without a vehicle; after removing origins with a missing input, they cover 120,400 origin--facility pairs from 890 CBGs. In these models, the estimated rural difference in modeled driving time falls from 108.5\% to 17.4\% after adding population density and measured demographics, and to 11.2\% after also adding potential access and tract-level household vehicle availability. The remaining association does not isolate facility scarcity from other unmeasured rural conditions. Road distance follows the same pattern. These outcomes describe geographic separation for represented trips; they do not include time spent reaching a vehicle or transit stop, waiting for a ride, or arranging travel around work and caregiving, and should be read as two measurable parts of travel burden rather than a complete measure of it.

\begin{table*}[t]
\centering
\small
\caption{Estimated rural difference in observed trip distance and modeled driving time. Values are percentage differences, with 95\% confidence intervals in parentheses. All models include county indicators and weight each origin--facility pair by its panel-observed visitor count.}
\label{tab:travel_decomposition}
\begin{tabular}{lccc}
\toprule
Outcome & Rurality and county & Add density and demographics & Add access and vehicle availability \\
\midrule
Straight-line distance & 95.4 (62.1, 135.6) & 11.5 ($-1.9$, 26.6) & 7.0 ($-0.7$, 15.2) \\
Road-network distance & 102.7 (67.6, 145.0) & 15.4 (3.2, 29.2) & 10.0 (2.4, 18.1) \\
Modeled driving time & 108.5 (73.4, 150.7) & 17.4 (3.3, 33.4) & 11.2 (2.5, 20.7) \\
\bottomrule
\end{tabular}
\end{table*}

\begin{figure}[t]
  \centering
  \begin{subfigure}[t]{0.73\linewidth}
    \centering
    \includegraphics[width=\linewidth]{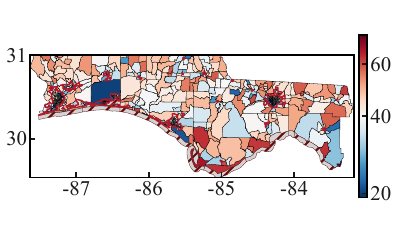}
    \Description{A Panhandle Census Block Group map uses a red-to-blue scale centered at age 40 to show median age, with urban-area boundaries outlined in red.}
    \caption{Age Median (Center: 40).}
  \end{subfigure}
  \hfill
  \begin{subfigure}[t]{0.24\linewidth}
    \centering
    \includegraphics[width=\linewidth]{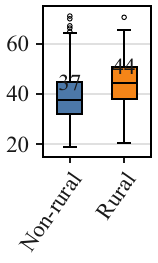}
    \Description{Box plots compare the distribution of Census Block Group median age in rural and non-rural Panhandle communities.}
    \caption{Age Median (CBG).}
  \end{subfigure}
  \caption{Median age of rural and non-rural Panhandle CBGs.}
  \label{fig:age_difference}
\end{figure}

The burden is compounded by demographic need (Figure~\ref{fig:age_difference}). The median age of rural CBGs is approximately seven years higher than that of non-rural CBGs. Older communities may require more frequent care and may face greater mobility constraints. The result therefore identifies a mismatch: communities likely to have elevated healthcare needs also face lower local facility availability and longer observed travel. This is an ecological association, not evidence that every older resident experiences the same barrier. In a visitor-weighted joint model with county fixed effects, rural origin is associated with 96\% longer modeled time (95\% CI 65\%--133\%) after including median age, income, uninsured share, race and ethnicity, and broadband. Median age has an adjusted association of $-2.1\%$ per standard deviation (95\% CI $-6.2\%$--2.1\%, $p=.318$).

Rural CBGs also have lower educational attainment in the descriptive ACS summaries. Prior research links directly measured health literacy with healthcare use and outcomes~\cite{berkman2011low}, but educational attainment is not equivalent to individual health literacy. We therefore use education only as demographic context and include income, broadband, disability, poverty, and household vehicle availability where the geographic resolution of the ACS data permits.

\subsection{RQ3: Temporal Dynamics and Disruption Response}
The number of facilities observed in the POI data remains relatively stable from January 2018 through June 2020 and declines thereafter. At every point, non-rural areas contain roughly four to five times as many observed facilities as rural areas. This persistent separation indicates that the supply disparity predates the pandemic and remains during the disruption period. Here, a decline denotes fewer facilities observed in the Patterns data, rather than independently verified closure or service loss. Figure~\ref{fig:appendix_facility_presence} plots the complete monthly series.

\begin{figure}[t]
  \centering
  \begin{subfigure}[t]{0.49\linewidth}
    \centering
    \includegraphics[width=\linewidth]{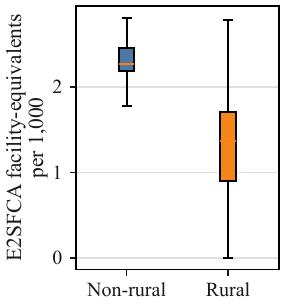}
    \Description{A box plot compares E2SFCA potential access for non-rural and rural Census Block Groups.}
    \caption{Potential spatial access.}
  \end{subfigure}
  \hfill
  \begin{subfigure}[t]{0.49\linewidth}
    \centering
    \includegraphics[width=\linewidth]{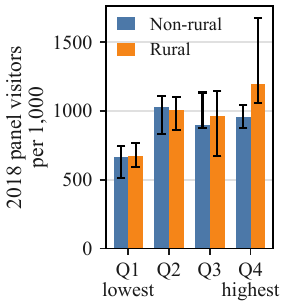}
    \Description{Grouped bars compare rural and non-rural panel visitor rates across four E2SFCA access quartiles.}
    \caption{Observed visitors across access quartiles.}
  \end{subfigure}
  \caption{Potential access and realized origin activity.}
  \label{fig:e2sfca}
\end{figure}

In 2018, rural origins account for 829 panel visitors per 1,000 residents, compared with 945 for non-rural origins. This difference does not imply lower healthcare need: residents who cannot reach a facility cannot generate an observed interaction regardless of need. We therefore compare realized activity with the E2SFCA potential-access measure developed in Section~\ref{sec:action_profiles} (Figure~\ref{fig:e2sfca}). Median potential access is 1.36 facility-equivalents per 1,000 in rural CBGs and 2.27 in non-rural CBGs; median modeled time to the nearest facility is 7.0 and 1.0 minutes, respectively. In an adjusted CBG model, E2SFCA access is positively associated with the log visitor rate ($p=.003$). The rural-by-access interaction is $-0.072$ (95\% CI $-0.185$--0.042, $p=.216$), so the data do not show that the rural visitor-rate gap consistently narrows as potential access improves.

\begin{figure*}[t]
  \centering
  \begin{subfigure}[t]{0.24\textwidth}
    \centering
    \includegraphics[width=\linewidth]{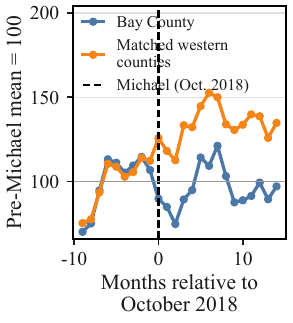}
    \Description{An event-time line chart compares Bay County destination visits with matched western counties around October 2018.}
    \caption{Visits to healthcare POIs: matched comparison.}
  \end{subfigure}
  \hfill
  \begin{subfigure}[t]{0.24\textwidth}
    \centering
    \includegraphics[width=\linewidth]{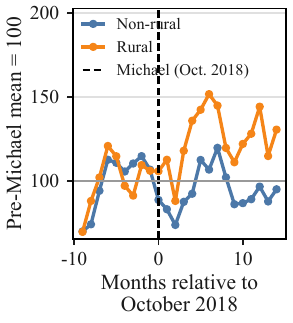}
    \Description{An event-time line chart compares Bay County rural and non-rural destination visits around October 2018.}
    \caption{Visits to healthcare POIs: Bay County by rurality.}
  \end{subfigure}
  \hfill
  \begin{subfigure}[t]{0.24\textwidth}
    \centering
    \includegraphics[width=\linewidth]{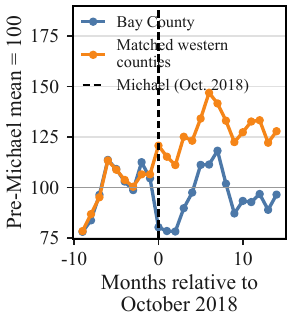}
    \Description{An event-time line chart compares Bay County origin panel visitors with matched western counties around October 2018.}
    \caption{Origin panel visitors: matched comparison.}
  \end{subfigure}
  \hfill
  \begin{subfigure}[t]{0.24\textwidth}
    \centering
    \includegraphics[width=\linewidth]{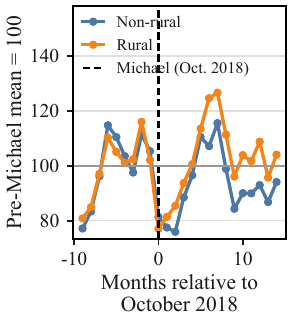}
    \Description{An event-time line chart compares Bay County rural and non-rural origin panel visitors around October 2018.}
    \caption{Origin panel visitors: Bay County by rurality.}
  \end{subfigure}
  \caption{Event-time paths around Hurricane Michael. Western donor weights are selected on January--September 2018; all series are indexed to their own pre-event mean. ``Origin'' denotes panel-observed visitors by home CBG, not latent demand.}
  \label{fig:hurricane_michael}
\end{figure*}

The COVID-19 period shows a shared decline in both rurality groups. Relative to the January--February 2020 baseline, April destination visits per observed facility fall by 52.6\% at non-rural facilities and 46.3\% at rural facilities; origin panel visitors per 1,000 residents fall by 58.7\% and 57.3\%, respectively. 

\subsection{Hurricane Michael Case Study}
Bay County provides a focused view of system response to Hurricane Michael (Figure~\ref{fig:hurricane_michael}). To distinguish the October 2018 change from the broader monthly pattern, we compare Bay County with Escambia and Santa Rosa Counties after normalizing each series to its January--September 2018 mean and selecting nonnegative donor weights over that pre-event period. Pre-event root-mean-square error is 3.31 index points for destination visits and 2.47 for origin visitors. In October 2018, Bay destination visits are 89.7 while the matched comparison is 125.8, a $-36.1$-point gap; the October--December average gap is $-35.7$. For origin visitors, the corresponding October and three-month gaps are $-40.4$ and $-36.7$ points. Within Bay County, October origin indices fall to 81.4 for non-rural CBGs and 77.1 for rural CBGs.

The divergence is consistent with hurricane-related disruption. The matched gap does not return within 10 index points through December 2019, indicating a persistent difference between Bay County and the western comparison. The nine-month pre-period and two-county donor pool, however, do not separate facility damage and road disruption from evacuation, population displacement, panel change, and remaining seasonal differences.


\subsection{COVID-19 Case Study}

The COVID-19 analysis separates three questions that can look similar in aggregate data: how many facilities appear in the monthly product, how much activity occurs, and how much activity is associated with residents of each origin CBG. The first is called facility presence, the second destination activity, and the third origin activity. A reduction in any of these measures does not by itself establish a facility closure or a reduction in healthcare need.

\begin{figure}[t]
  \centering
  \includegraphics[width=\linewidth]{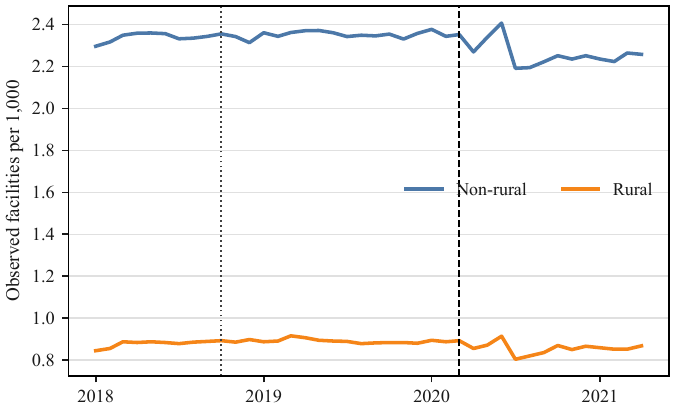}
  \Description{Monthly lines show the number of facilities observed per 1,000 residents in rural and non-rural Panhandle Census Block Groups from January 2018 through April 2021.}
  \caption{Monthly facility presence in the mobility product. The dotted vertical line marks Hurricane Michael in October 2018, and the dashed line marks March 2020.}
  \label{fig:appendix_facility_presence}
\end{figure}

Figure~\ref{fig:appendix_facility_presence} shows the number of facilities observed each month per 1,000 residents. Both rural and non-rural series are fairly stable through June 2020 and lower afterward. The rural series remains below the non-rural series throughout the study period. Because presence depends on whether a facility appears in the monthly Patterns product, the later decline may combine real operational changes with changes in product coverage.

Activity falls much more sharply than facility presence. In April 2020, destination visits per observed facility are 52.6\% below the January--February 2020 mean in non-rural areas and 46.3\% below it in rural areas, whereas origin panel visitors per 1,000 residents are 58.7\% and 57.3\% lower. The rural and non-rural paths therefore move largely in parallel, and the pre-existing gap in local supply persists through the disruption. An emergency can increase healthcare need without increasing recorded visits when elective care is postponed, residents evacuate, facilities restrict entry, or encounters move to channels not represented in the data. 

\section{RQ4: From Spatial Evidence to Community Action}
RQ1--RQ3 identify where local healthcare supply is limited, which communities experience the greatest travel and demographic burden, and how observed activity changes during disruptions. RQ4 translates this evidence into a four-step decision workflow: (1) quantify community-specific access deficits, (2) prioritize communities transparently, (3) match barriers to feasible interventions and optimize deployment, and (4) evaluate operational and equity outcomes. The framework supports rather than replaces the decisions of healthcare providers, transportation agencies, emergency managers, and community organizations.

Applying a transparent screen to the Panhandle identifies 45 rural CBGs containing 53,612 residents that are in the lowest E2SFCA access quartile, lack a locally observed ambulatory facility, and are more than 15 modeled minutes from the nearest facility. Twenty-seven of these CBGs, containing 28,026 residents, also have a median age at or above the rural CBG median of 43.9 years. The 15-minute threshold is the first travel band in the E2SFCA calculation. The flagged CBGs span 16 Panhandle counties, and the older subset spans 14, so these communities are geographically dispersed rather than concentrated in one county: Walton and Jackson each contain six flagged CBGs, and Jefferson and Santa Rosa each contain four.

This result supports immediate planning without pretending that the available data determine a clinic route. The flagged CBGs provide a focused set of communities for checking candidate mobile-clinic sites, transportation availability, desired services, and feasible operating times with local partners. The older subset may require accessible vehicles, shorter waiting times, and closer coordination with chronic-care providers. A deployment model can be solved after partners specify candidate stops, staffing and capacity, operating costs, the number of mobile units, and acceptable travel times.

\subsection{Step 1: Build Community Access Profiles}
\label{sec:action_profiles}
Per-capita facility counts reveal broad disparities but do not account for cross-boundary access, competition for services, or travel impedance. Instead of applying a simple adjacency-based spatial lag, we use an enhanced two-step floating catchment area (E2SFCA) score to capture cross-boundary access through road-network travel time, distance decay, and competition for services. Let $S_j$ denote the service capacity of facility $j$, $P_k$ the demand population in CBG $k$, $\tau_{kj}$ the road-network travel time from $k$ to $j$, and $f(\tau)$ a distance-decay function. The provider-to-demand ratio for facility $j$ is
\begin{equation}
R_j=\frac{S_j}{\sum_{k:\tau_{kj}\leq \tau_0}P_k f(\tau_{kj})},
\end{equation}
and the potential accessibility of CBG $i$ is
\begin{equation}
A_i^{\mathrm{pot}}=\sum_{j:\tau_{ij}\leq \tau_0}R_j f(\tau_{ij}).
\end{equation}
The catchment threshold $\tau_0$ and decay function may be adjusted to the travel range that residents consider acceptable. In the empirical analysis, $\tau_0=60$ minutes, the four decay weights are 1.00, 0.68, 0.22, and 0.05, and $S_j=1$ for every facility observed in 2018 because provider capacity is unavailable.

Potential access is complemented by realized travel burden:
\begin{equation}
B_i^{\mathrm{real}}=\frac{\sum_j v_{ij}\tau_{ij}}{\sum_j v_{ij}},
\end{equation}
where $v_{ij}$ is the panel-observed visitor allocation from CBG $i$ to facility $j$. This measure is undefined for CBGs with no represented panel visitors; these CBGs retain their potential-access and local-supply measures and are flagged as missing for realized travel. Let $A_i^{\star}$ denote a partner-defined potential-access benchmark and define the access deficit as $G_i^{\mathrm{pot}}=\max(0,A_i^{\star}-A_i^{\mathrm{pot}})$. Each community is then represented by an interpretable barrier profile,
\begin{equation}
\mathbf{h}_i=[G_i^{\mathrm{pot}},\ B_i^{\mathrm{real}},\ \mathbb{I}(F_i=0),\ O_i,\ T_i,\ D_i],
\end{equation}
where $O_i$ captures age- or disability-related need, $T_i$ transportation disadvantage, and $D_i$ disruption vulnerability. The components should be reported separately so that trade-offs remain visible.

\subsection{Step 2: Prioritize Communities Transparently}
When planners require a single ranking, the profile can be summarized using a stakeholder-weighted multi-criteria score:
\vspace{-5pt}
\begin{equation}
Q_i=\sum_{k=1}^{K}w_k z(h_{ik}), \qquad w_k\geq0,\quad \sum_k w_k=1,
\end{equation}
where $z(\cdot)$ standardizes each indicator. Weights can be elicited through the analytic hierarchy process, swing weighting, or structured partner workshops. The analysis should report rank stability across plausible weights and identify communities whose priority is sensitive to stakeholder assumptions. This avoids presenting one arbitrary weighting scheme as an objective ground truth.

\subsection{Step 3: Match Barriers to Interventions}
Different access profiles call for different interventions. Table~\ref{tab:action_mapping} maps representative patterns to technical formulations and decision outputs. The mapping is not automatic: local feasibility, resident preference, staffing, operating budgets, and partner-defined constraints determine the final intervention.

\begin{table*}[t]
\centering
\small
\caption{Mapping spatial evidence to candidate community interventions and technical methods.}
\label{tab:action_mapping}
\begin{tabular}{p{0.22\textwidth}p{0.15\textwidth}p{0.29\textwidth}p{0.25\textwidth}}
\toprule
Community access pattern & Candidate intervention & Technical formulation & Decision output \\
\midrule
No local ambulatory facility, low E2SFCA access, and a large older population & Mobile clinic or rotating service & Maximal covering location, capacitated $p$-median, or location-routing optimization & Weekly stops, service frequency, assigned CBGs, and expected coverage \\
Nearby regional capacity but long trips and low vehicle availability & Non-emergency medical transportation & Dial-a-ride or vehicle routing with time windows and accessibility constraints & Pickup schedules, pooled routes, fleet requirements, and expected ride time \\
High ambulatory travel burden and adequate broadband readiness & Telehealth access points and digital support & Capacitated facility location or hub allocation & Priority sites, staffing levels, and assigned communities \\
High hurricane exposure and strong event-related disruption & Mobile resilience hubs and backup services & Two-stage stochastic or robust optimization across disruption scenarios & Pre-positioned resources, backup sites, and scenario-specific redeployment \\
\bottomrule
\end{tabular}
\end{table*}

\paragraph{Technical example: mobile-clinic deployment.}
The access screen can support a future mobile-clinic planning exercise. Let $\mathcal{I}$ denote priority CBGs, $\mathcal{J}$ candidate stops such as partner-approved public sites, and $\mathcal{T}$ weekly service periods. Let $q_i$ be priority-weighted demand derived from $Q_i$,
$\tau_{ij}$ the road-network travel time, $a_{ij}=1$ when $\tau_{ij}$ is within an acceptable threshold, $y_{jt}$ indicate whether a clinic operates at site $j$ in period $t$, and $x_{ijt}$ indicate whether demand from CBG $i$ is assigned to that stop. One formulation is
\begin{align}
\max_{x,y}\quad & \sum_{i,j,t}q_i x_{ijt}
-\lambda\sum_{i,j,t}q_i\tau_{ij}x_{ijt}
-\mu\sum_{j,t}c_{jt}y_{jt} \\
\text{s.t.}\quad
& \sum_j x_{ijt}\leq 1, && \forall i,t,\\
& x_{ijt}\leq a_{ij}y_{jt}, && \forall i,j,t,\\
& \sum_i q_i x_{ijt}\leq C_{jt}y_{jt}, && \forall j,t,\\
& \sum_j y_{jt}\leq K_t, && \forall t,\\
& x_{ijt},y_{jt}\in\{0,1\},
\end{align}
where $C_{jt}$ is service capacity, $K_t$ is the number of available clinic units or stops, and $c_{jt}$ is operating cost. The objective maximizes priority demand served while penalizing resident travel and deployment cost. Equity can be imposed through minimum-coverage constraints for the highest-need CBGs or a penalty on the worst-group access gap.
For example, a rural CBG with no ambulatory facility, low E2SFCA access, an older population, and more than 30 minutes of road travel may be assigned to a weekly stop at a partner-approved public site together with nearby communities. Lower-priority CBGs can share the stop when residual capacity is available, reducing deployment cost without displacing the highest-need community. Once candidate sites and operating constraints are available, a solved model would determine where the clinic should stop, how often it should operate, which CBGs it should serve, and how many residents would gain access within a policy-relevant travel threshold.

\subsection{Step 4: Co-Design and Evaluate Deployment}
Technical plans need to be validated with community partners. Partners should identify feasible stops, operating hours, service categories, capacity limits, acceptable travel thresholds, and unacceptable solutions. Resident input can reveal barriers not captured in spatial data, including caregiving schedules, distrust, language needs, or inaccessible pickup locations. Partner input should be documented as concrete changes to weights, candidate sites, constraints, or evaluation criteria, rather than treated only as dissemination.
A deployed intervention can be evaluated through a phased rollout or matched before--after design. For outcome $Y_{it}$ in community $i$ and month $t$, a difference-in-differences specification is
\begin{equation}
Y_{it}=\alpha_i+\gamma_t+\beta(\mathrm{Treated}_i\times\mathrm{Post}_t)+\boldsymbol{\theta}^{\top}\mathbf{X}_{it}+\epsilon_{it},
\end{equation}
where $\beta$ estimates the change associated with deployment under the design assumptions. The evaluation will also estimate event-time coefficients and plot pre-treatment coefficients and confidence intervals so that the parallel-trends assumption can be assessed. $\mathbf{X}_{it}$ will include prespecified time-varying insurance-policy changes, concurrent health or transportation interventions, panel coverage, seasonal storm activity, and other documented shocks. Outcomes may include completed visits, missed appointments, median travel time, utilization in priority CBGs, cost per completed visit, resident satisfaction, and recovery time after disruption. Equity should be assessed through worst-group improvement and changes in access gaps across rurality, age, income, and other partner-identified groups. Because deployment has not yet been completed, we specify the optimization component as a retrospective planning design and regard the partner-confirmed pilot protocol as future work.

\section{Limitations and Future Work}
\textbf{Privacy and ethical data use.} The study uses aggregated facility-visit data provided by Dewey and does not require identifying individual trajectories for the reported analyses. No attempt was made to identify individuals, infer sensitive diagnoses, or publish sparse origin--destination flows that could expose residents or facilities. Future validation with individual-level mobility or clinical records could rely on privacy-preserving synthetic data, such as generated activity traces or longitudinal electronic health records~\cite{xu2026synhat,li2026synehr}, before access to protected data is arranged.

\textbf{Representativeness.} Smartphone-derived mobility data may underrepresent older adults, low-income residents, people with limited broadband or smartphone access, and sparsely populated rural communities~\cite{li2023bias,brelsford2022publicspace,jay2020income,grantz2020mobile,kang2020multiscale}. These characteristics overlap with the barriers studied here. Missing high-barrier residents may understate a disparity, whereas geographic differences in panel retention or privacy suppression may overstate one. Population normalization adjusts for population size but not panel composition. Florida discharge or all-payer claims data would provide an important external validation source but are not available in the present study.

\textbf{Measurement limitations.} The pooled POI inventory is not a regulatory facility census, and the 2018-observed sensitivity depends on Patterns coverage. Facility presence is not identical to operating capacity, provider staffing, service quality, appointment availability, or affordability. Observed activity captures behavior within the panel, not unmet need, foregone care, telehealth, or every healthcare interaction. Road times use representative points and road-class speed assumptions rather than observed routes or congestion. They measure geographic separation under a motor-vehicle assumption, not total door-to-door burden; vehicle access, public transit, shared rides, walking, and actual travel mode are not observed. 


\textbf{Unobserved confounding.} Rurality, facility siting, socioeconomic composition, road access, and panel coverage are mutually related and none is randomly assigned. The adjusted supply and travel models account for measured differences but cannot remove omitted variables such as land cost, local regulation, provider quality, or health status. The E2SFCA and utilization comparisons remain descriptive, and the Michael analysis provides a comparison path rather than a causal storm-effect estimate.

\textbf{Temporal and causal limitations.} Changes around Hurricane Michael and COVID-19 may reflect seasonality, evacuation, panel composition, economic disruption, or telehealth substitution. The mobility data end in April 2021. Lower observed activity alone does not demonstrate lower need, financial distress, or facility closure without administrative and financial validation.

\section{Conclusion}
This study uses multi-source spatial data to characterize rural healthcare accessibility and resilience in Florida. The results show a consistent mismatch between rural geography and healthcare access: rural communities occupy most of the region's land area, nearly half of rural CBGs contain no facility in the pooled inventory, and rural residents have less than half the per-capita facility availability of non-rural residents. The supply disparity remains after measured demographic adjustment. Among origin--facility pairs represented in the mobility panel, rural road distance and modeled travel time are more than twice the non-rural averages, although population density explains much of the difference. Rural CBGs also have lower E2SFCA potential access. Longitudinal results show a persistent divergence in Bay County after Hurricane Michael and broad activity declines in both rural and non-rural communities during the early COVID-19 period.
The central implication is that rural healthcare planning should not rely on facility counts or telehealth alone. Spatial evidence can focus stakeholder decisions about transportation, mobile care, broadband, health communication, and disaster resilience on specific communities while operational constraints are collected for deployment planning.

\begin{acks}
We thank all the reviewers for their insightful feedback and comments. Dr. Wang is partially supported by the Florida State University (FSU) Startup Fund and the Institute for Successful Longevity (ISL) planning grant. Dr. He is partially supported by the National Institute of Mental Health grant R33MH137736, and the University of Florida-Florida State University Clinical and Translational Science Award UM1TR005128.
\end{acks}

\bibliographystyle{ACM-Reference-Format}
\bibliography{references}

\end{document}